\documentclass[%
reprint,
showpacs,
showkeys,
nofootinbib,
nobibnotes,
amsmath,fixed
amssymb,
aps,
superscriptaddress
]{revtex4-1}

\usepackage[english]{babel}
\usepackage[utf8x]{inputenc}
\usepackage[T1]{fontenc}
\usepackage[left]{lineno}

\usepackage{amsmath}
\usepackage{graphicx}
\usepackage[export]{adjustbox}

\usepackage[colorlinks=true,linkcolor=blue,citecolor=blue,urlcolor=blue]{hyperref}
\usepackage{epstopdf}
\usepackage{amssymb}
\usepackage{color}
\usepackage{xcolor}
\usepackage{soul}
\usepackage{url}
\usepackage[justification=justified]{subcaption}
\usepackage[justification=justified, figurename=FIG.]{caption}

\usepackage{float}
\restylefloat*{figure}
\usepackage{placeins}

\usepackage{dcolumn}
\usepackage{bm} 

\usepackage{siunitx}
\usepackage{textcomp} 
\usepackage{multirow}
\usepackage{tabularx} 
\usepackage{ragged2e}
\usepackage[overload]{textcase}

\begin{document}
\title{Search for New Physics via Low-Energy Electron Recoils with a 4.2 Tonne$\times$Year Exposure from the LZ Experiment}

\date{\today}

\author{D.S.~Akerib}
\affiliation{SLAC National Accelerator Laboratory, Menlo Park, CA 94025-7015, USA}
\affiliation{Kavli Institute for Particle Astrophysics and Cosmology, Stanford University, Stanford, CA  94305-4085 USA}

\author{A.K.~Al Musalhi}
\affiliation{University College London (UCL), Department of Physics and Astronomy, London WC1E 6BT, UK}

\author{F.~Alder}
\affiliation{University College London (UCL), Department of Physics and Astronomy, London WC1E 6BT, UK}

\author{J.~Almquist}
\affiliation{Brown University, Department of Physics, Providence, RI 02912-9037, USA}

\author{C.S.~Amarasinghe}
\affiliation{University of California, Santa Barbara, Department of Physics, Santa Barbara, CA 93106-9530, USA}

\author{A.~Ames}
\affiliation{SLAC National Accelerator Laboratory, Menlo Park, CA 94025-7015, USA}
\affiliation{Kavli Institute for Particle Astrophysics and Cosmology, Stanford University, Stanford, CA  94305-4085 USA}

\author{T.J.~Anderson}
\affiliation{SLAC National Accelerator Laboratory, Menlo Park, CA 94025-7015, USA}
\affiliation{Kavli Institute for Particle Astrophysics and Cosmology, Stanford University, Stanford, CA  94305-4085 USA}

\author{N.~Angelides}
\affiliation{Imperial College London, Physics Department, Blackett Laboratory, London SW7 2AZ, UK}

\author{H.M.~Ara\'{u}jo}
\affiliation{Imperial College London, Physics Department, Blackett Laboratory, London SW7 2AZ, UK}

\author{J.E.~Armstrong}
\affiliation{University of Maryland, Department of Physics, College Park, MD 20742-4111, USA}

% ...

\collaboration{The LZ Collaboration}

\begin{abstract}
\noindent
We report results from searches for new physics models through electron recoils using data collected by the LUX-ZEPLIN (LZ) experiment during its first two science runs, with a total exposure of 4.2 tonne-years. 
The observed data are consistent with a background-only hypothesis. 
Constraints are derived for electromagnetic interactions of solar neutrinos, solar axion-like particles (ALPs), mirror dark matter, and the absorption of bosonic dark matter candidates.
The inverse Primakoff process for $^{57}$Fe de-excitation solar ALPs is considered for the first time.
\textcolor{black}{
These results represent the most stringent constraints to date on keV-scale Primakoff and $^{57}$Fe solar ALPs, bosonic dark matter, mirror dark matter,  and neutrino millicharge, while remaining competitive for the other signal models investigated.
}
\end{abstract}


\maketitle
Dual-phase xenon time projection chambers (TPCs), combining a low energy threshold, scalability to large volumes, and ultra-low background rates \cite{LZ:2022ysc}, have become a cornerstone technology for experimental searches for rare processes, particularly for the detection of weakly interacting massive particles (WIMPs) \cite{LZ:2024zvo, XENON:2025vwd, PandaX:2024qfu}, \textcolor{black}{a well-motivated} dark matter (DM) candidate.
While these detectors have traditionally focused on nuclear recoil (NR) signals expected from dark matter interactions \cite{LZ:2022lsv,  LZ:2024psa, LZ:2024vge, LZ:2024trd, LZ:2024zvo,  LZ:2025iaw, PandaX-4T:2021bab, PandaX:2024qfu, XENON:2023cxc, XENON:2025vwd}, they also provide excellent sensitivity to electronic recoil (ER) signals \cite{LZ:2023poo, XENON:2022ltv, PandaX:2024cic,LZ:2025xkj}.
This opens avenues to a broad spectrum of new physics searches other than conventional WIMPs.
\textcolor{black}{In this Letter, we report on searches for new physics scenarios that produce ER signals, using data collected by the LUX-ZEPLIN (LZ) experiment. 
This analysis probes additional new physics models compared with previous LZ ER studies~\cite{LZ:2023poo} and is the first to apply the radon tagging technique and to adopt a signal-dependent analysis and statistical treatment for low-energy electron-recoil searches.
}

Axions are pseudoscalar Nambu--Goldstone bosons arising from the Peccei--Quinn mechanism introduced to solve the strong CP problem~\cite{Peccei:1977hh}. 
Axion-like particles (ALPs) share similar couplings to Standard Model (SM) particles, \textcolor{black}{but do not in general satisfy the mass--coupling relation characteristic of QCD axions~\cite{Cirelli:2024ssz, Carenza:2024qaq}.
With effective tree-level couplings to photons, electrons, or nucleons, ALPs can be produced in the hot solar-core plasma through thermally driven processes, with characteristic energies set by the keV-scale solar temperature.}
We consider solar ALPs produced in the Primakoff conversion induced by the photon coupling $g_{a\gamma\gamma}$\textcolor{black}{~with dimension GeV$^{-1}$}~\cite{Raffelt:1996wa, Caputo:2024oqc}, ABC processes (atomic recombination and de-excitation, bremsstrahlung, and Compton scattering) resulting from the \textcolor{black}{dimensionless} electron coupling $g_{ae}$~\cite{Dimopoulos:1986kc, Dimopoulos:1986mi, Pospelov:2008jk, Raffelt:1985nk, Zhitnitsky:1979cn, Krauss:1984gm, Mikaelian:1978jg, Fukugita:1982ep, Fukugita:1982gn}, and $^{57}$Fe nuclear de-excitation through the \textcolor{black}{dimensionless} effective nucleon coupling $g^\text{eff}_{aN}~$, enabled by its low-lying 14.4 keV excited state that can be thermally populated in the solar core~\cite{Serenelli:2009yc,Moriyama:1995bz,Haxton:1991pu}.
For Primakoff ALPs, we use the mass-dependent solar flux of Ref.~\cite{Wu:2024fsf} and the inverse Primakoff scattering cross section from the relativistic Hartree--Fock treatment of Ref.~\cite{Abe:2020mcs}. \textcolor{black}{Unlike previous experimental studies, this search retains the mass dependence in both the flux and the cross section model.
For $^{57}$Fe ALPs, we use the mass-dependent event rate from
Refs.~\cite{CAST:2009jdc, Armengaud:2013rta, PandaX:2024cic}, which is
phase-space suppressed as $m_a$ approaches the excited state energy and vanishes above it. 
The deposited-energy spectrum of $^{57}$Fe ALPs is modeled as a monoenergetic line at
14.4~keV.}
We consider both axio-electric and inverse Primakoff interactions in xenon, the latter being explored experimentally here for the first time for solar 
$^{57}$Fe ALPs.
For more details of the solar ALP modeling, please see Sec.~I of the Supplemental Material~\cite{SupplementalMaterial}.
\nocite{Redondo:2013wwa,Derbin:2012yk,Benito:2020lgu,bienayme2024dark}

We search for bosonic halo dark matter in the keV/c$^2$ mass range \cite{Pospelov:2008jk}.
Bosonic dark matter may be produced as pseudoscalars (e.g., ALPs produced via the early-universe misalignment mechanism \cite{Preskill:1982cy, Adams:2022pbo}) or as vectors (hidden photons (HP) \cite{Fabbrichesi:2020wbt, Cline:2024qzv}), generated through similar dynamics \cite{Nelson:2011sf}.
In such cases, bosonic DM may get absorbed by atoms in the liquid xenon (LXe) bulk, producing ER-like monoenergetic line signals via processes analogous to the photoelectric effect~\cite{Pospelov:2008jk, PandaX:2024cic, XENON:2020rca}.
For both cases, the absorption rate on electrons is evaluated using calculations in Refs.~\cite{Pospelov:2008jk, Bloch:2016sjj}, assuming a local dark-matter density of 0.3 GeV/cm$^3$ \cite{deSalas:2019pee}.
\textcolor{black}{We note that the results depend on the assumed local density and leave detailed discussion of the modeling in Sec.~II of the Supplemental Material~\cite{SupplementalMaterial}.}

We study the mirror dark matter (MDM) model, which represents a theoretically motivated framework for plasma dark matter~\cite{Clarke:2016eac} by postulating an entire dark sector isomorphic to the SM~\cite{Foot:2014mia}.
In this model, the dark sector interacts with SM matter via gravity and photon-mirror photon mixing~\cite{Foot:1991bp}, which will lead to mirror electrons scattering with electrons in LXe~\cite{Clarke:2016eac, LUX:2019gwa}.
In this search, the signal is modeled following Ref.~\cite{Foot:2018jpo}, with the local mirror-helium density taken to be $5.8\times10^{-11}$ cm$^{-3}$, consistent with Ref.~\cite{LUX:2019gwa}.
\textcolor{black}{We also include the small time-averaged annual/diurnal modulation of the mirror-electron event rate predicted in this model. When averaged over the live time, this changes the expected event rate by about 2\%.
For more details of MDM, refer to Sec.~III of the Supplemental Material~\cite{SupplementalMaterial}.
}

The last search in this analysis is the electromagnetic interaction of solar neutrinos.
\textcolor{black}{Neutrinos do not couple to photons at tree level.
In the minimally extended SM, with loop-level correction applied, the Dirac magnetic moment is only of order $10^{-20} \mu_B$ for $\mathcal{O}(0.1)$ eV neutrino masses, while the neutrino millicharge remains zero~\cite{Fujikawa:1980yx, Schechter:1981hw}.
Such contributions are far below current detection sensitivity, so any observable neutrino electromagnetic signal would imply new physics, with a magnetic-moment interpretation additionally bearing on the Dirac/Majorana nature of the neutrino~\cite{Bell:2005kz, Bell:2006wi}.
}
For such interactions from solar neutrinos, we use the flux from the standard solar model \cite{Vinyoles:2016djt} and the cross section from Ref.~\cite{Giunti:2014ixa} with corrections from the random-phase approximation \cite{Hsieh:2019hug} applied.
\textcolor{black}{
More details are provided in Sec.~IV of the Supplemental Material~\cite{SupplementalMaterial}.
}

\textcolor{black}{
The data used in this search were collected during the first two science runs of the LZ experiment: WIMP search 2022 (WS2022) \cite{LZ:2022lsv} and WIMP search 2024 (WS2024) \cite{LZ:2024zvo}, lasting from 2021 to 2024.
The LZ experiment is located 4850 feet underground at the Sanford Underground Research Facility (SURF) in Lead, South Dakota, USA. 
At the center of the experiment is a dual-phase xenon TPC containing 7 tonnes of active xenon~\cite{LZ:2019sgr} as the target medium. 
To suppress radioactive backgrounds, the TPC is surrounded by a 2-tonne LXe skin detector and a 17.3-tonne gadolinium-loaded liquid scintillator outer detector for $\gamma$ and neutron tagging, all within a 238-tonne ultra-pure water shield~\cite{LZ:2019sgr}.
}

\textcolor{black}{
Particle interactions in the TPC produce prompt scintillation (S1) and ionization electrons; electrons drifting to the liquid--gas interface under an electric field are extracted into the gas phase and generate secondary electroluminescence (S2)~\cite{LZ:2022lsv}. The S1--S2 time difference and S2 light pattern provide 3D position reconstruction, while the S2/S1 ratio enables ER/NR discrimination~\cite{Dahl:2009nta}. Spatial and temporal variations in S1 and S2 are mapped with periodically injected monoenergetic ER sources, including $^{131\text{m}}$Xe and $^{83\text{m}}$Kr~\cite{LZ:2024bsz}, defining corrected observables S1c and S2c; both are expressed in detected photons (phd), accounting for double-photoelectron emission~\cite{Faham:2015kqa,LopezParedes:2018kzu}.
}
\\
\indent 
Calibration campaigns are performed to study the detector's response to energy deposits from particle interactions.
Specifically, the detector's response to ER events is calibrated using $\beta$ decays of tritiated methane (CH$_3$T), $^{14}$C~\cite{LZ:2022lsv, LZ:2024zvo} and $^{220}$Rn~\cite{LZ:2023lvz}.
The scintillation ($g_1$) and ionization ($g_2$) gains are obtained by tuning simulations generated with the Noble Element Simulation Technique (\textsc{nest})~\cite{Szydagis:2024nest} to match calibration data, with additional constraints from monoenergetic peaks.
The best-fit results are $g_1 = 0.114 \pm 0.002$~phd/photon and $g_2 = 47.1 \pm 1.1$~phd/electron for WS2022~\cite{LZ:2022lsv}, and $g_1 = 0.112 \pm 0.002$~phd/photon and $g_2 = 34.0 \pm 0.9$~phd/electron for WS2024~\cite{LZ:2024zvo}. 
\textcolor{black}{
The different $g_2$ values mainly reflect the different operating conditions of WS2022 and WS2024, especially the extraction field. 
As ER sensitivity depends weakly on ER/NR discrimination, this configuration change does not significantly impact LZ's ER sensitivity.}
\\
\indent 
The dominant contributor to the ER backgrounds in both LZ science runs is the $\beta$ decay of $^{214}$Pb with no coincident $\gamma$ emissions. 
This isotope originates from $^{222}$Rn, which emanates from detector materials and is dispersed into the liquid xenon bulk, resulting in a measured fiducial volume activity of 3.9$\pm$0.4 $\mu$Bq/kg \cite{LZ:2024zvo}. 
To mitigate this background, a novel radon-tagging technique was developed to identify events with a high likelihood of being a $^{214}$Pb $\beta$ decay, and applied to the WS2024 dataset~\cite{McLaughlin:2024syp, LZ:2025xxf}. 
This is achieved by creating a slow and laminar xenon flow pattern in the central region of the TPC by adjusting the liquid xenon circulation rate, cooling power, and temperature gradients.
Then the liquid velocity is mapped using coincident $^{222}$Rn-$^{218}$Po $\alpha$ decay pairs, enabling tracking of the movement of daughter atoms following the $^{218}$Po alpha decay.
In practice, a tag volume comoving with the  $^{218}$Po daughter is defined and lasts for three times the $^{214}$Pb half-life (27 minutes), in which it is probable to find a $^{214}$Pb $\beta$ decay.
Single scatter (SS) events that occur within this volume and time window contribute to a $^{214}$Pb-rich dataset, containing 15$\%$ of the WS2024 exposure and approximately 60$\pm$4\% of the total $^{214}$Pb activity \cite{LZ:2024zvo}. 
The rest of the data, not being tagged as $^{214}$Pb, forms a dataset with a lower effective $^{214}$Pb activity of 1.8$\pm$0.3~$\mu$Bq/kg. 
As part of the likelihood, the radon-tagged events are also included in the final statistical analysis as a sideband dataset to improve the ability to constrain the tagging efficiency and other background contributions.
The tagging efficiency is left to float as a nuisance parameter that scales the event rates in both the $^{214}$Pb-rich and $^{214}$Pb-poor datasets.
\textcolor{black}{Since $^{214}$Pb is the dominant ER background, this subdivision gives the likelihood additional leverage to constrain the leading background component and thereby improves the sensitivity to ER signals.}
\textcolor{black}{
Xenon electron capture (EC) and double electron capture (DEC) decays also contribute to the ER background with smaller S2/S1 ratio than $\beta$ decays because their higher ionization density enhances recombination. 
Measurements with neutron-activated $^{127}$Xe and $^{125}$Xe give the ratio between charge yields of L-shell decay and tritium $\beta$ decay, $Q_{y}^{\mathrm{L}}/Q_{y}^{\beta} = 0.87 \pm 0.03$ \cite{LZ:2025hud}, confirming suppressed charge yield in EC. 
This suppression is incorporated in the background model. 
For $^{124}$Xe DEC, LM captures (5.98 keV) use a fixed factor of 0.87, while LL captures (10.00 keV) are described by a Thomas--Imel-motivated parameter floated in the fit~\cite{LZ:2024zvo,Temples:2021jym,LZ:2025hud,Xu:2025ksk}.
Other ER backgrounds include dissolved $\beta$ emitters in the LXe bulk, detector-material $\gamma$ Compton scatters, solar neutrino--electron scatters, cosmogenic $^{37}$Ar with a 35-day half-life at the beginning of WS2022, and residual CH$_3$T and $^{14}$C from calibrations in WS2024~\cite{LZ:2022kml,LZ:2022ysc}.
Additional subdominant backgrounds from solar and atmospheric CE$\nu$NS, detector NRs, and accidental coincidences are modeled as in Refs.~\cite{LZ:2022lsv,LZ:2024zvo} and have negligible impact on the final result, especially for 1D absorption-like analyses.
}

\textcolor{black}{
This analysis uses 280 live days of data, with a total exposure of 4.2 $\pm$ 0.1 tonne-years, of which 0.9 tonne-years are from WS2022 and the other 3.3 tonne-years from WS2024.}
A fiducial volume is defined for each science run to suppress external backgrounds, 
and the fiducial mass for both runs is $5.5 \pm 0.2$~tonnes~\cite{LZ:2022lsv, LZ:2024zvo}.
The drift field was reduced from 193 V/cm in WS2022 to 97 V/cm in WS2024 to mitigate light emission in the skin detector.
Data selection criteria and acceptances are determined for each run using a combination of calibration datasets and simulations.
For this work, we retain the same data selection strategy as in Refs.~\cite{LZ:2022lsv, LZ:2024zvo}, except for the region of interest (ROI).

To reflect the characteristics of different signal models, two ROI definitions are considered in this work.
For all signal models, events are required to satisfy S1c > 3 phd in both science runs.
The lower bound on uncorrected S2 size is 600 phd for WS2022 and 645 phd for WS2024, corresponding to 10 and 14.5 extracted electrons, respectively.
The solar neutrino electromagnetic interaction and mirror dark matter scattering are expected to create pure $\beta$-like recoils in LXe due to their scattering nature, similar to solar neutrino weak interactions with electrons in LXe~\cite{LZ:2021xov, LZ:2023poo}. 
The recoil energy spectra of those events scale as  $\mathrm{d}\Gamma/\mathrm{d}E_r \sim 1/E_r$ or $ 1/E_r^2$ because of the massless mediator, with most of the events concentrated near the energy threshold.
As those signals are nearly fully contained in the ROI of both the WS2022 and WS2024 analyses, and the detector response to $\beta$-like recoils is well understood through CH$_3$T calibrations, we follow Refs.~\cite{LZ:2022lsv,LZ:2024zvo} and set the ROI upper bound as S1c < 80 phd, and perform the analysis in the 2D (S1c, S2c) observable space for maximal consistency.
We will refer to this ROI as the 2D ROI henceforth.

\begin{figure}[ht]
    \centering
    \includegraphics[width=\columnwidth]{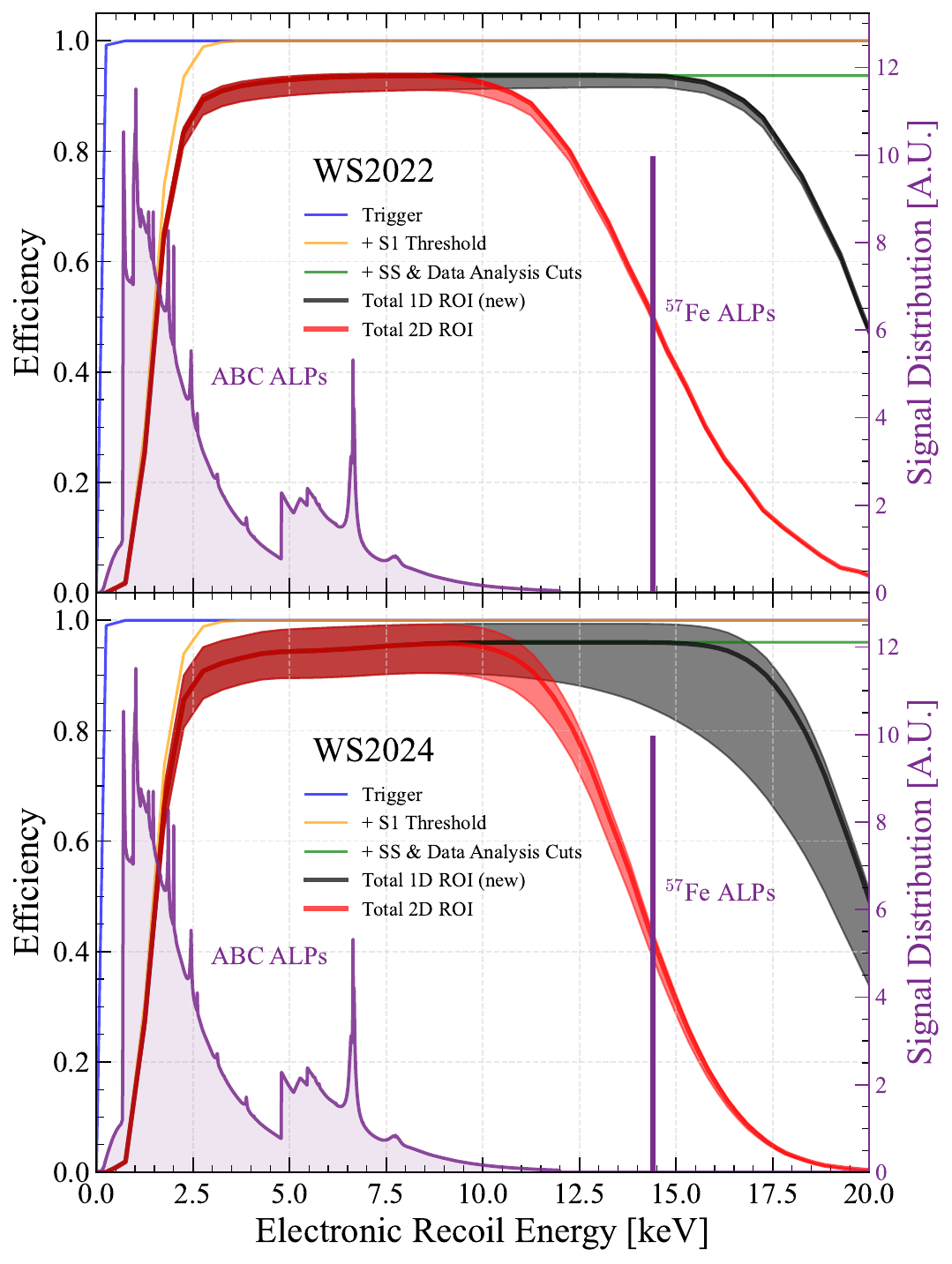}
    \caption{The total data selection efficiencies of the 1D ROI (black) and 2D ROI (red), with uncertainties (shaded), in WS2022 and WS2024 data.
    \textcolor{black}{The 1D ROI is introduced in this work.}
    The trigger (blue), three-fold coincidence and > 3 phd on S1c (yellow), single-scatter reconstruction and data analysis cuts (green) efficiencies are also shown.
    The 1D ROI exhibits a slower efficiency drop at higher energies, which can be attributed to replacing the upper S1c bound of 80 phd with a requirement of $E_\text{rec} < 20$ keV.
    \textcolor{black}{The true recoil energy distributions of solar ABC ALP and $^{57}$Fe ALP events are also shown for comparison.}
    }
    \label{fig:SR3Efficiency}
\end{figure}

For solar ALPs and bosonic DM, detection in LXe relies either on absorption of an exotic particle analogous to the photoelectric effect~\cite{Pospelov:2008jk}, or the photoelectric effect of the inverse Primakoff photon.
Previous studies found that photoelectric absorptions, albeit conserving total energy, display a different partition between light and charge yields~\cite{Szydagis:2024nest}.
As there is no pure photon calibration available \textcolor{black}{below 20 keV}, it is difficult to construct a reliable signal model for such processes in the traditional (S1c, S2c) space.
Given this limitation, we model signal and background and conduct statistical inference in the reconstructed energy $E_\text{rec}$ space, following the treatment in earlier experimental studies \cite{XENON:2022ltv, PandaX:2024cic}.
The $^{57}$Fe signals are at $14.4$ keV, at which energy the WIMP analysis selection efficiency has begun to roll off due to the S1c < 80 phd bound of the 2D ROI.
To increase signal efficiency for $^{57}$Fe and heavy bosonic DM signals, we replace the S1c upper bound with an $E_\text{rec} < 20$ keV boundary.
In practice, this is implemented through requiring $W_q(\tfrac{\text{S1c}}{g_1} +  \tfrac{\text{S2c}}{g_2})$ < 20 keV, where {$W_q=13.5$~eV~\cite{Szydagis:2021hfh, Dahl:2009nta, Goetzke:2016lfg} is the work function of xenon.
Hereafter, we refer to this as the 1D ROI.
\begin{table*}[]
    \caption{\justifying \color{black} Background-only fit results for the WS2022 (top) and WS2024 (bottom) datasets, for both the 1D and 2D ROIs. The WS2024 ER backgrounds are grouped according to how their rates are affected by radon tagging.
    In the 1D ROI introduced in this work, the detector NR and neutrino CE$\nu$NS backgrounds are not included due to negligible rates and lack of discrimination.   
    }
    \centering
    \setlength{\tabcolsep}{14pt} 

\begin{tabular}{p{5mm} l r@{}l r@{}l r@{}l r@{}l}
    \hline
    \hline
    \noalign{\vskip 1mm}
    \multicolumn{2}{c}{} & 
    \multicolumn{4}{c}{\textbf{1D ROI (new)}} & 
    \multicolumn{4}{c}{\textbf{2D ROI}} \\
    \hline
    \multicolumn{2}{c}{\textbf{Source}} & 
    \multicolumn{2}{c}{\textbf{Expectation}} & 
    \multicolumn{2}{c}{\textbf{Fit result}} & 
    \multicolumn{2}{c}{\textbf{Expectation}} & 
    \multicolumn{2}{c}{\textbf{Fit result}} \\
    \colrule \noalign{\vskip 1mm}

    {\multirow{8}{*}{\rotatebox[origin=c]{90}{\textbf{WS2022}}}}
    &$\beta$ decays + Detector $\gamma$s  & 305 &~$\pm$ 50 & 336 &~$\pm$ 20 & 215 &~$\pm$ 36 & 222 &~$\pm$ 16 \\
    &Solar $\nu$ ER  & 38.0 &~$\pm$ 2.3 & 38.1 &~$\pm$ 2.3 & 27.1 &~$\pm$ 1.6 & 27.2 &~$\pm$ 1.6 \\
    &$^{136}$Xe 2$\nu\beta\beta$ & 27.2 &~$\pm$ 4.4 & 27.8 &~$\pm$ 4.3 & 15.1 &~$\pm$ 2.4 & 15.2 &~$\pm$ 2.4 \\
    &$^{127}$Xe + $^{125}$Xe EC & 9.2 &~$\pm$ 0.8 & 9.3 &~$\pm$ 0.8 & 9.2 &~$\pm$ 0.8 & 9.3 &~$\pm$ 0.8 \\
    &$^{124}$Xe DEC & 5.0 &~$\pm$ 1.4 & 5.0 &~$\pm$ 1.4 & 5.0 &~$\pm$ 1.4 & 5.2 &~$\pm$ 1.4 \\
    &$^{37}$Ar  & [0,&~288] & 47.6 &~$\pm$ 9.4 & [0,&~288] & 52.7 &~$\pm$ 9.4 \\
    &Accidental coincidences  & 1.2 &~$\pm$ 0.3 & 1.2 &~$\pm$ 0.3 & 1.2 &~$\pm$ 0.3 & 1.2 &~$\pm$ 0.3 \\
    &Det NR & -& & - & & -&  & 0.07 &~$^{+0.20}_{-0.07}$ \\
    &Solar $^8$B NR & -& & -& & 0.14&$~\pm$ 0.01 & 0.14 &~$\pm$ 0.01 \\

    \hline
    &Total  & -&  & 464 &~$\pm$ 23 & -&  & 333 &~$\pm$ 19 \\
    \colrule \noalign{\vskip 0.75 mm}
    \colrule \noalign{\vskip 0.75 mm}


    {\multirow{13}{*}{\rotatebox[origin=c]{90}{\textbf{WS2024}}}}
    &$^{214}$Pb $\beta$  & 1099 &~$\pm$ 130 & 1096 &~$\pm$ 46 & 743 &~$\pm$ 88 & 733 &~$\pm$ 34 \\
    &$^{85}$Kr+$^{39}$Ar+$\beta$s+Detector $\gamma$s  & 237 &~$\pm$ 33 & 236 &~$\pm$ 31 & 162 &~$\pm$ 22 & 161 &~$\pm$ 21 \\
    &Solar $\nu$ ER  & 151 &~$\pm$ 9.0 & 151 &~$\pm$ 9.0 & 102 &~$\pm$ 6.1 & 102 &~$\pm$ 6.1 \\
    &$^{136}$Xe 2$\nu\beta\beta$ & 107 &~$\pm$ 16 & 108 &~$\pm$ 16 & 55.6 &~$\pm$ 8.4 & 55.8 &~$\pm$ 8.2 \\
    &$^{218}$Po + $^{212}$Pb $\beta$s  & 92.1 &~$\pm$ 11.0 & 93.1 &~$\pm$ 10.9 & 62.7 &~$\pm$ 7.5 & 63.6 &~$\pm$ 7.4 \\
    &$^{3}$H+$^{14}$C  & 61.2 &~$\pm$ 3.4 & 62.9 &~$\pm$ 3.4 & 58.3 &~$\pm$ 3.3 & 59.7 &~$\pm$ 3.3 \\
    &$^{124}$Xe DEC & 20.0 &~$\pm$ 4.0 & 19.8 &~$\pm$ 4.0 & 19.4 &~$\pm$ 3.9 & 21.4 &~$\pm$ 3.6 \\
    &$^{127}$Xe + $^{125}$Xe EC   & 3.2 &~$\pm$ 0.7 & 2.8 &~$\pm$ 0.6 & 3.2 &~$\pm$ 0.7 & 2.7 &~$\pm$ 0.6 \\
    &Atmospheric $\nu$ NR & -& & - & & 0.12&$~\pm $ 0.02 & 0.12 &~$\pm$ 0.03 \\
    &Solar $^8$B + hep $\nu$ NR & -& & - & & 0.06&$~\pm $ 0.01 & 0.06 &~$\pm$ 0.01 \\

    &Det NR & -& & - & & -& & 0.0 &~$^{+0.2}$ \\

&Rn $\beta$ tagging efficiency & 0.60 &$~\pm$~0.04 & 0.62 &~$\pm$ 0.02 & 0.60&$~\pm~$0.04 & 0.62 &~$\pm$ 0.02 \\
    \hline
    &Total  & 1773&~$\pm~$136 & 1754&~$\pm$~60  & 1210 &~$\pm$ 91  & 1202 &~$\pm$ 42 \\
    \hline
    \hline
    \end{tabular}
    \label{tab:counts}
\end{table*}

\begin{figure}[ht!]
    \centering
    \includegraphics[width=\columnwidth]{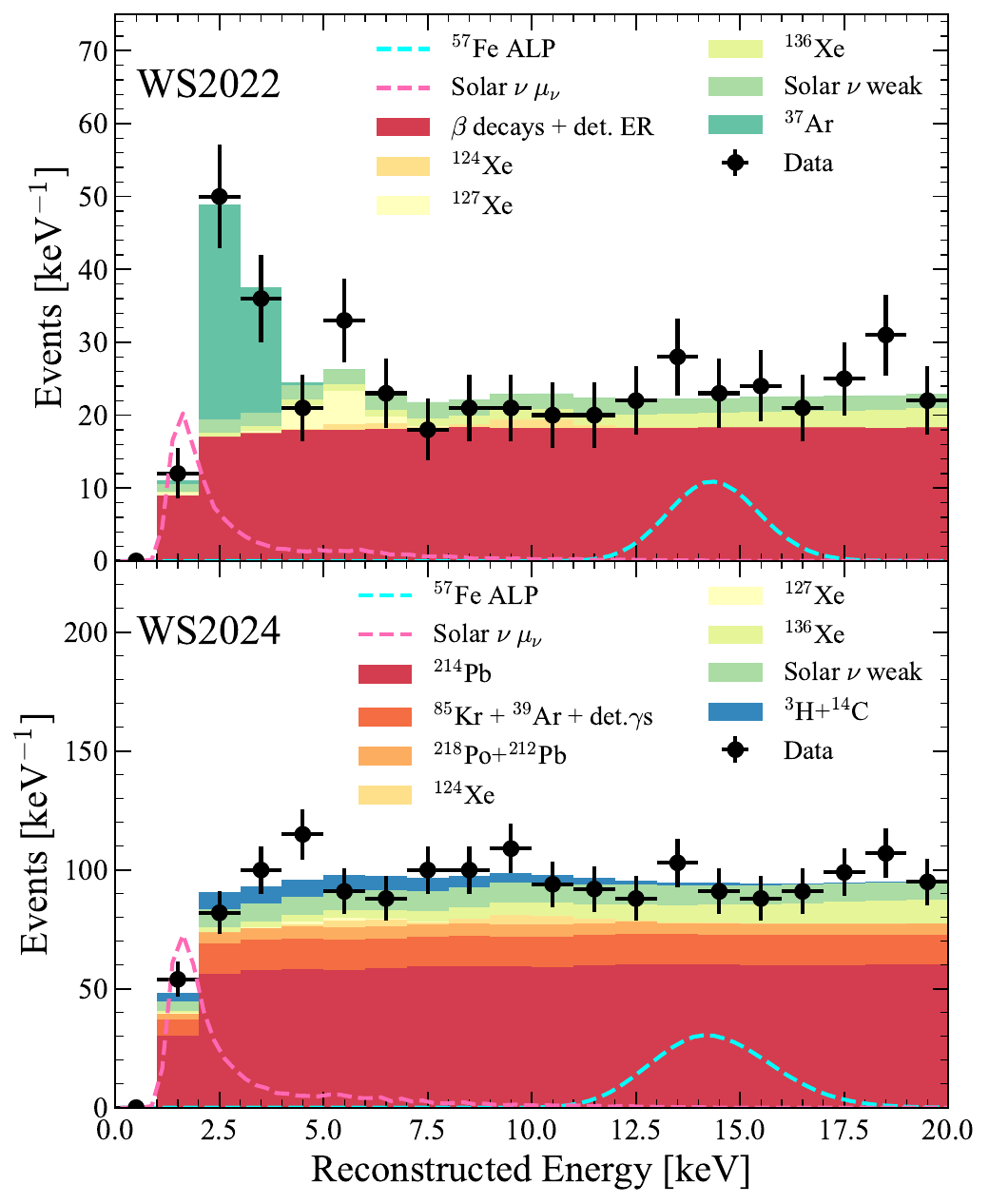}
    \caption{
    LZ WS2022  + 2024 data (black dots) and best-fit background-only model (colored histogram) using the 1D likelihood.
    \textcolor{black}{The region above 15 keV is reported publicly for the first time in this work. 
    Simulated spectra for two representative signals, solar neutrino magnetic moment with $\mu_\nu = 2.8\times10^{-11}\mu_B$ (magenta) and $^{57}$Fe solar ALPs with $m_a = 1~\mathrm{meV}/c^2$ and $g_{ae}g_{aN}^{\mathrm{eff}} = 6.1\times10^{-18}$ (cyan), are overlaid for comparison. A fit in the WIMP search 2D ROI gives a very similar background-only description and is presented in Ref.~\cite{LZ:2024zvo}.}}
    \label{fig:dataPlot}
\end{figure}

\indent The trigger efficiency, SS reconstruction efficiency, and analysis-cut acceptances are evaluated using calibration data \textcolor{black}{together with simulations validated against those data. 
The impact of the ROI definitions, particularly the high-energy roll-off, is modeled using the same tuned simulations.}
Figure~\ref{fig:SR3Efficiency} illustrates the cumulative impact of different data selection stages on the ER signal efficiency as a function of true energy in both ROI settings, for WS2022 and WS2024, respectively. 
The uncertainty is evaluated from the statistical uncertainty and the difference between datasets, as in Refs.~\cite{LZ:2022lsv,LZ:2024zvo}. 
The true energy spectra for each physics signal and background component, except for accidentals, are passed through the LZ parametric simulation chain \cite{LZSims} based on  \textsc{nest}-2.4.0 \cite{Szydagis:2024nest} to create simulation datasets.
The data selection efficiencies are applied to simulations to create probability distribution functions (PDFs) in the desired observable space as inputs for statistical inference.


To mitigate analyzer bias, LZ injected synthetic “salt” events at random times into the dataset during data taking.
\textcolor{black}{All the salt events in these runs were designed for WIMP searches, hence NR-like; no salt was introduced specifically for this ER search.
While the region corresponding to the 2D ROI had already been unsalted for the analysis presented in Ref.~\cite{LZ:2024zvo}, the energy-only 1D ROI used here extends outside that previously unsalted region and could therefore still contain higher-energy NR-like salt. 
The region between 1D and 2D ROI bounds was kept salted during the development of this work and was unsalted only after all analysis choices were frozen. 
No additional salt events were found.}

\textcolor{black}{Both the 1D and 2D background-only fits show good agreement between the post-unsalting data and the background model, with goodness-of-fit \(p\)-values of 0.80 and 0.24, respectively.}
The 2D analysis incorporates detector neutron and neutrino CE$\nu$NS backgrounds, which in the 1D case can be effectively absorbed into the ER background sources due to their small event rates.
The yield parameter for LL captures is allowed to vary in the 2D fits but fixed in the 1D case, since charge suppression has little impact on the 1D analysis.
The pre- and post-fit event rates of different background components considered in the 1D and 2D fits are given in Table~\ref{tab:counts}.

As the background-only fits show good compatibility with observation, we follow conventions set in \cite{Baxter:2021pqo} to set upper limits for the signal models using a two-sided profile likelihood ratio (PLR) approach.
\textcolor{black}{
The dataset after unsalting} and best fit to the background-only model are shown in Fig.~\ref{fig:dataPlot}.
The PLR test statistics~\cite{Cowan:2010js} are evaluated using energy-only 1D PDFs for the solar ALPs and bosonic dark matter absorptions, while 2D PDFs in (S1c, S2c) are used for the mirror DM and solar neutrino electromagnetic interactions.
The sensitivity and \textcolor{black}{power-constrained} observed upper limits of models tested are shown in Figs.~\ref{fig:1DResults} and \ref{fig:2DResults}.

\begin{figure*}[t]
    \centering
    \includegraphics[width=\textwidth]{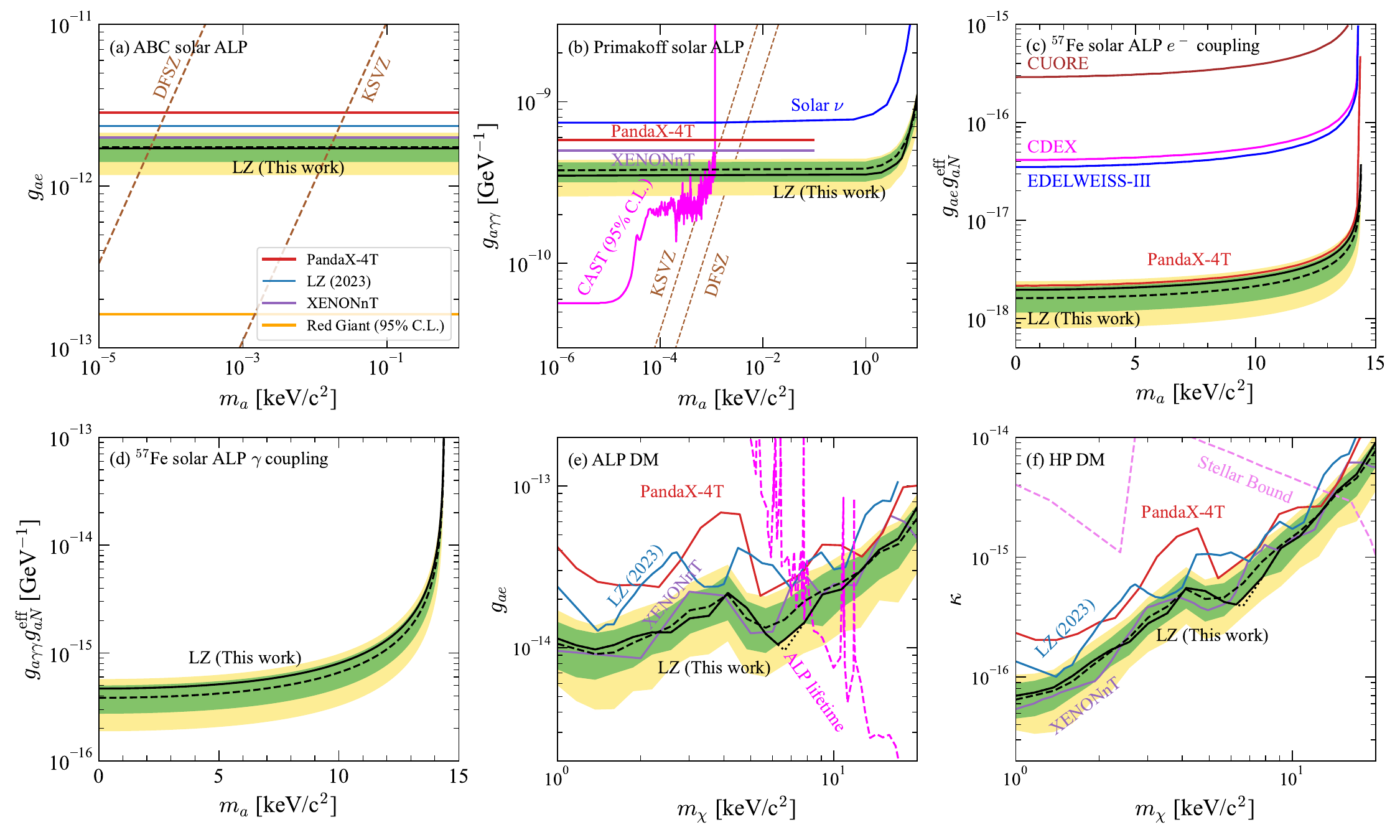}
    \caption{The sensitivities and \textcolor{black}{power-constrained} 90\% C.L. limits on new physics models tested using 1D statistical inference: 
    (a) ABC solar ALP, (b) Primakoff solar ALP, (c) $^{57}$Fe solar ALP electron coupling, (d) $^{57}$Fe solar ALP photon coupling, (e) ALP DM, and (f) \textcolor{black}{hidden photon} (HP) DM.
    \textcolor{black}{
    For the LZ results, the solid black curves show the reported observed power-constrained $90\%$ C.L. upper limits, the dashed black curves the median expected sensitivities, and the dotted black curves in panels (e) and (f) the upper limits obtained without applying the power constraint. The green (yellow) bands indicate the central $68\%$ ($95\%$) ranges of the expected sensitivities under the background-only hypothesis.}
    Selected constraints from terrestrial experiments~\cite{LZ:2023poo, XENON:2022ltv, PandaX:2024cic, CAST:2009jdc, CUORE:2012ymr, CDEX:2019exx, EDELWEISS:2018tde} and astrophysical observations~\cite{Ferreira:2022egk, An:2014twa, Capozzi:2020cbu, Vinyoles:2015aba}, together with the regions allowed by two benchmark QCD axion realizations, KSVZ~\cite{Kim:1979if} and DFSZ~\cite{Dine:1981rt}, are also shown for comparison. 
    }
    \label{fig:1DResults}
\end{figure*}
\begin{figure*}[t]
    \centering
    \includegraphics[width=\textwidth]{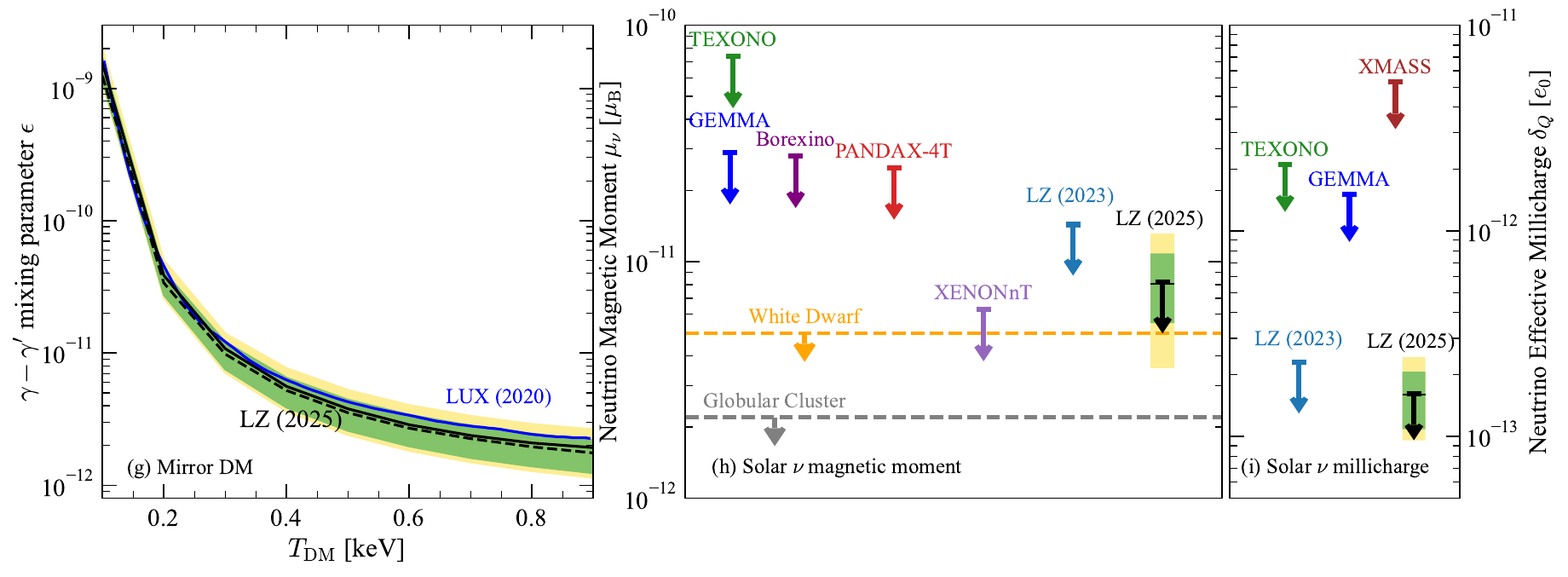}
    \caption{The sensitivities and 90\% C.L. limits on the new physics models tested using 2D statistical inference: (g) mirror dark matter, (h) solar neutrino magnetic moment and (i) millicharge. Constraints from terrestrial experiments~\cite{LUX:2019gwa, TEXONO:2009knm, Beda:2009kx,Borexino:2017fbd, XENON:2020rca, XENON:2022ltv, XMASS:2020zke} and astrophysical observations~\cite{Corsico:2014mpa, Viaux:2013lha} are also shown.  
    \textcolor{black}{
    The black curves and colored bands follow the same plotting conventions as in Fig.~\ref{fig:1DResults}.
    }
    }
    \label{fig:2DResults}
\end{figure*}

This analysis sets stringent constraints across multiple physics searches. 
\textcolor{black}{We obtain an upper limit of $g_{ae}<1.8\times10^{-12}$ for the dimensionless ALP-electron coupling,} which is a 15\% improvement over leading terrestrial constraints \cite{PandaX:2024cic, XENON:2022ltv}. 
\textcolor{black}{The limit on ALP-photon coupling is $g_{a\gamma\gamma}<3.5\times10^{-10}$ GeV$^{-1}$, providing the most stringent constraints to date for ALPs with masses of $\sim1$ eV/c$^2$ and above, and improving upon previous leading constraints by 30\% \cite{PandaX:2024cic, XENON:2022ltv}.}
Our sensitivity extends above $m_a =1 $~keV/c$^2$ thanks to the newly introduced mass-dependent Primakoff flux \cite{Wu:2024fsf} and cross section \cite{Abe:2020mcs}.
The upper limit on $^{57}$Fe de-excitation ALP event rate in LXe is $14$ events tonne$^{-1}$~year$^{-1}$, \textcolor{black}{a 30$\%$ improvement over previous constraints on $g_{ae}g_{aN}^\text{eff}$~\cite{PandaX:2024cic, XENON:2022ltv}.
This leads to the most stringent constraints on $g_{ae}g_{aN}^\text{eff}$, and the first experimental constraints on $g_{a\gamma\gamma}g_{aN}^\text{eff}$.}

For solar neutrino electromagnetic interactions, we set the magnetic moment constraint on $\mu_\nu < 8.3 \times 10^{-12}\,\mu_B$, a factor of two stronger than our WS2022 result~\cite{LZ:2023poo}. 
\textcolor{black}{This improvement is driven primarily by the larger combined exposure of the WS2022+WS2024 dataset and introduction of the radon-tagging technique.}
We also set the current leading limit on the neutrino millicharge, $\delta_Q < 1.6\times10^{-13}~e_0$.
The mirror dark matter temperature above $T_\mathrm{DM}$ = 0.29 keV is excluded; in terms of kinetic mixing, we set $\epsilon < 4\times10^{-12}$ for the photon--mirror-photon kinetic-mixing parameter at $T_{\rm DM}=0.5$ keV, which surpasses the previous best \cite{LUX:2019gwa} by 10\%. 
Lastly, our bosonic DM absorption limits are comparable with the leading constraints reported in Ref.~\cite{XENON:2022ltv}, complementary to the astrophysical constraints on ALP-electron coupling derived from X-ray and $\gamma$-ray observations~\cite{Ferreira:2022egk} and the constraints on the mixing strength between hidden photons and photons from stellar bounds~\cite{An:2014twa}.

\textcolor{black}{Overall, the sensitivity of this analysis is slightly stronger than those of XENONnT and PandaX-4T \cite{XENON:2022ltv, PandaX:2024cic}.
While this search benefits from a larger total exposure and radon tagging, the higher overall ER background rate than XENONnT partially offsets these advantages.
The relative strength of the observed limits does not follow a uniform pattern across models, but instead depends on the signal's spectral shape and, consequently, on fluctuations in the local data in the relevant energy region.
}

The introduction of radon tagging has improved LZ's sensitivity to ER signals.
Fits to toy datasets indicate that incorporating radon tagging reduces the median upper limits on signal events by approximately 10--25$\%$, \textcolor{black}{and the median sensitivities to a similar extent}.
\textcolor{black}{
The improvement is smaller than the tagging efficiency because the tagging efficiency carries a systematic uncertainty, originating mainly from the $^{214}$Pb branching ratio~\cite{LZ:2025xxf}.} As a result, the upper limits derived from observed data typically improve by about 5--10$\%$, with the exact gain depending on the signal's spectral shape and underlying physics, \textcolor{black}{particularly the exponent with which the reported model parameter enters the expected signal rate.}
The only exception is the $^{57}$Fe solar ALPs, for which the observed upper limit with radon tagging is slightly weaker. This is consistent with a modest local upward fluctuation in the reconstructed-energy spectrum around 14.4 keV, \textcolor{black}{visible in Fig. 2 as the data lie slightly above the background-only model in the region where the $^{57}$Fe signal template peaks.}

\textcolor{black}{More importantly, this work demonstrates that xenon TPCs are powerful probes of new physics in the ER channel.
With the development of radon background control techniques~\cite{LZ:2025xxf, XENON:2025nic} and the scaling up of exposure, their reach for keV-scale electron-coupling new physics will continue to improve.
Based on next-generation baseline designs~\cite{XLZD:2024nsu} and the Asimov median significance defined in Ref.~\cite{Cowan:2010js}, we expect the sensitivity to improve by $\mathcal{O}(10)$ in terms of event rate, leading to a factor of $\sim3$ improvement for models whose rates are quadratic in the coupling and $\sim 2$ for models with quartic coupling dependence. 
A more detailed sensitivity calculation for next-generation xenon-based experiments is left to future work.
Nevertheless, with these advancements, significant new parameter space for solar ALPs and MDM can be explored, and the sensitivity to the neutrino magnetic moment will become comparable to current astrophysical constraints.
}

In conclusion, we have searched for ER signals indicative of new physics using the combined dataset from the first two science runs of the LZ experiment. 
Finding no significant excess above the background model, we set stringent limits on various signal hypotheses, thanks to a combination of low background rate, large exposure, and the introduction of the radon tagging technique. 
The results provide the strongest constraints on Primakoff and $^{57}$Fe de-excitation ALPs, solar $\nu$ millicharge and mirror dark matter, while the constraints on ABC solar ALPs, solar neutrino magnetic moment and bosonic dark matter absorption remain highly competitive.
\textcolor{black}{These results further establish low-energy ER searches in liquid xenon TPCs as a competitive avenue for probing new physics, and we expect further gains from larger exposures and improved background control.}
\enlargethispage{\baselineskip} 

\textit{Acknowledgements---}The research supporting this work took place in part at the Sanford Underground Research Facility (SURF) in Lead, South Dakota. Funding for this work is supported by the U.S. Department of Energy, Office of Science, Office of High Energy Physics under Contract Numbers DE-AC02-05CH11231, DE-SC0020216, DE-SC0012704, DE-SC0010010, DE-AC02-07CH11359, DE-SC0015910, DE-SC0014223, DE-SC0010813, DE-SC0009999, DE-NA0003180, DE-SC0011702, DE-SC0010072, DE-SC0006605, DE-SC0008475, DE-SC0019193, DE-FG02-10ER46709, UW PRJ82AJ, DE-SC0013542, DE-AC02-76SF00515, DE-SC0018982, DE-SC0019066, DE-SC0015535, DE-SC0019319, DE-SC0025629, DE-SC0024114, DE-AC52-07NA27344, \& DE-SC0012447. This research was also supported by U.S. National Science Foundation (NSF); the UKRI’s Science \& Technology Facilities Council under award numbers ST/W000490/1, ST/W000482/1, ST/W000636/1, ST/W000466/1, ST/W000628/1, ST/W000555/1, ST/W000547/1, ST/W00058X/1, ST/X508263/1, ST/V506862/1, ST/X508561/1, ST/V507040/1 , ST/W507787/1, ST/R003181/1, ST/R003181/2,  ST/W507957/1, ST/X005984/1, ST/X006050/1; Portuguese Foundation for Science and Technology (FCT) under award numbers PTDC/FIS-PAR/2831/2020; the Institute for Basic Science, Korea (budget number IBS-R016-D1); the Swiss National Science Foundation (SNSF)  under award number 10001549. This research was supported by the Australian Government through the Australian Research Council Centre of Excellence for Dark Matter Particle Physics under award number CE200100008. We acknowledge additional support from the UK Science \& Technology Facilities Council (STFC) for PhD studentships and the STFC Boulby Underground Laboratory in the U.K., the GridPP~\cite{GridPP:2006wnd, Britton:2009ser} and IRIS Collaborations, in particular at Imperial College London and additional support by the University College London (UCL) Cosmoparticle Initiative, and the University of Z\"urich. We acknowledge additional support from the Center for the Fundamental Physics of the Universe, Brown University. Y. Xu has received funding from the European Union's Horizon Europe research and innovation programme under the Marie Skłodowska-Curie grant agreement No. 101126636. K.T. Lesko acknowledges the support of Brasenose College and Oxford University. The LZ Collaboration acknowledges the key contributions of Dr. Sidney Cahn, Yale University, in the production of calibration sources. This research used resources of the National Energy Research Scientific Computing Center, a DOE Office of Science User Facility supported by the Office of Science of the U.S. Department of Energy under Contract No. DE-AC02-05CH11231. We gratefully acknowledge support from GitLab through its GitLab for Education Program. The University of Edinburgh is a charitable body, registered in Scotland, with the registration number SC005336. The assistance of SURF and its personnel in providing physical access and general logistical and technical support is acknowledged. We acknowledge the South Dakota Governor's office, the South Dakota Community Foundation, the South Dakota State University Foundation, and the University of South Dakota Foundation for use of xenon. We also acknowledge the University of Alabama for providing xenon. For the purpose of open access, the authors have applied a Creative Commons Attribution (CC BY) license to any Author Accepted Manuscript version arising from this submission. Finally, we respectfully acknowledge that we are on the traditional land of Indigenous American peoples and honor their rich cultural heritage and enduring contributions. Their deep connection to this land and their resilience and wisdom continue to inspire and enrich our community. We commit to learning from and supporting their effort as original stewards of this land and to preserve their cultures and rights for a more inclusive and sustainable future.

\textit{Appendix: Data availability---}Selected data from this analysis are publicly available \cite{Zenodo_record}, including the following:
\begin{itemize}
    \item Figure.~\ref{fig:SR3Efficiency}: the total efficiencies of data selection cuts for both ROIs and datasets.
    \item Figure.~\ref{fig:dataPlot}: \textcolor{black}{The 1D histograms and the data in reconstructed energy used to create those histograms.}
    \item Figure.~\ref{fig:1DResults},~\ref{fig:2DResults}: the \textcolor{black}{power-constrained} observed upper limits on signal models reported in this analysis.
\end{itemize}

\bibliographystyle{apsrev4-2}
\bibliography{main}

\end{document}